\documentclass[lengthcheck,superscriptaddress]{revtex4-1}
\usepackage{graphicx}
\usepackage{amsmath}
\usepackage{hyperref}
\usepackage{bm}
\usepackage[dvips]{color}

\graphicspath{{figs/},
      }

\newcommand{\fzd} {Helmholtz-Zentrum Dresden-Rossendorf, Institute of Ion Beam Physics and Materials Research, Bautzner Landstrasse 400
01328 Dresden, Germany}
\newcommand{\nottingham} {School of Physics and Astronomy, University of Nottingham, Nottingham NG7 2RD, United Kingdom}
\newcommand{\ios} {Institute of Semiconductors, Chinese Academy of Science, Beijing 100080, China}
\newcommand{\tud} {Institute of Applied Physics, Technische Universit\"{a}t Dresden, 01062 Dresden, Germany}
\newcommand{\tudt} {Institute of Theoretical Physics, Technische Universit\"{a}t Dresden, 01062 Dresden, Germany}
\newcommand{\buct} {Department of Physics and Electronics, School of Science, Beijing University of Chemical Technology, Beijing 100029, China}

\begin{document}

\title{Precise tuning of the Curie temperature of (Ga,Mn)As-based magnetic
semiconductors by hole compensation: Support for valence-band
ferromagnetism}
\date{\today}

\author{Shengqiang~Zhou}
\email[Electronic address: ]{S.Zhou@hzdr.de}
\affiliation{\fzd}
\author{Lin~Li}
\affiliation{\fzd} \affiliation{\buct}
\author{Ye~Yuan}
\affiliation{\fzd}
\author{A. W.~Rushforth}
\affiliation{\nottingham}
\author{Lin~Chen}
\affiliation{\ios}
\author{Yutian~Wang}
\affiliation{\fzd}
\author{R.~B\"{o}ttger}
\affiliation{\fzd}
\author{R.~Heller}
\affiliation{\fzd}
\author{Jianhua~Zhao}
\affiliation{\ios}
\author{K. W.~Edmonds}
\author{R. P.~Campion}
\affiliation{\nottingham}
\author{B. L.~Gallagher}
\affiliation{\nottingham}
\author{C.~Timm}
\affiliation{\tudt}
\author{M.~Helm}
\affiliation{\fzd}
\affiliation{\tud}

\begin{abstract}
For the prototype diluted ferromagnetic semiconductor
(Ga,Mn)As, there is a fundamental concern about the electronic
states near the Fermi level, i.e., whether the Fermi level resides
in a well-separated impurity band derived from Mn doping (impurity-band
model) or in the valence band that is already merged with the Mn-derived
impurity band (valence-band model). We investigate this
question by carefully shifting the Fermi level by means of carrier
compensation. We use helium-ion implantation, a standard industry
technology, to precisely compensate the hole doping of GaAs-based
diluted ferromagnetic semiconductors while keeping  the Mn concentration constant. We monitor the change
of Curie temperature ($T_C$) and conductivity. For a broad range of
samples including (Ga,Mn)As and (Ga,Mn)(As,P) with various Mn and
P concentrations, we observe a smooth decrease of $T_C$ with carrier
compensation over a wide temperature range while the conduction is changed from
metallic to insulating. The existence of $T_C$ below 10\,K is also confirmed in
heavily compensated samples. Our experimental results
are naturally explained within the valence-band picture.
\end{abstract}

\maketitle

%%%%%%%%%%%%%%%%%%%%%%%%%%%Introduction%%%%%%%%%%%%%%%%%%%%%%%%%%%%%%%%%%%%

\section{Introduction}

As one of three routes to spintronics \cite{RevModPhys.86.187},
diluted ferromagnetic semiconductors (DFS) have been under
extensive investigation for two decades
\cite{RevModPhys.78.809,Dietl_10y,Timm2012,bouzerar2015unraveling}. Mn-doped
GaAs [and its alloys such as Ga(As,P) and Ga(As,Sb)] is regarded as a
prototype. Spintronic devices
\cite{ohnoSPINLED,chiba03,ciorga2007tamr,ciorga2009electrical}, hybrid
structures \cite{stolichnov08,Korenev12}, and other spintronic
phenomena \cite{vila2007universal,endres2013demonstration,gorchon2014stochastic}
have been successfully demonstrated. One
of the bottlenecks limiting its application is the low Curie
temperature $T_C$, which still does not reach room
temperature~\cite{chenlin2011}.

Different theoretical models suggest distinct strategies for increasing
$T_C$. In its extreme limit, the impurity-band picture assumes the
existence of a separate impurity band formed by Mn-derived dangling-bond
hybrid (DBH) states \cite{Zunger1986,PhysRevB.68.075202,Timm2012}. In the
presence of compensation, the Fermi energy falls into this impurity band. This
picture is expected to be correct for weakly doped samples.
On the other hand, the extreme valence-band picture proposes that the
would-be impurity band is so strongly broadened that it merges with the valence
band and does not leave a distinct maximum in the
density of states (DOS). The only effect of Mn doping, besides moving the Fermi
energy into the merged band, is to create a tail of localized states at the band
edge \cite{RevModPhys.78.809,Timm2012}. If compensation is not too strong, the
states at the Fermi energy are extended and have valence-band character with
some admixture of DBHs.
There is a continuum of cases connecting these two extremes and some of
the controversy
\cite{PhysRevLett.97.087208,PhysRevB.73.024411,PhysRevB.76.125206,PhysRevLett.89.097203,PhysRevB.84.081203,
PhysRevLett.103.137201,PhysRevLett.105.227201,Sawichi2010,Ohya2011,
Dobrowolska2012,Samarth2012} in the field seems to result from overinterpreting
the two pictures. In this work, we use the term ``impurity-band picture'' to
refer to a situation where the impurity band may have merged with the valence
band but is still visible by a clear maximum in the DOS. This means that in
uncompensated systems, the Fermi energy falls into a minimum of the DOS, where
the electronic states show an enhanced tendency towards localization. The low
DOS and the localization generically lead to reduced conductivity and magnetic
coupling. Hence, within the impurity-band model $T_C$ is expected to decrease
as compensation goes to zero, whereas partial compensation of holes promotes
ferromagnetic order and $T_C$ reaches its maximum when the impurity band is
roughly half filled \cite{PhysRevLett.89.227201,Dobrowolska2012,PhysRevB.87.205314}.
Within the valence-band picture, one instead expects the DOS and $T_C$ to
increase monotonically with hole concentration. In the extreme limit,
Dietl \textit{et al.}\ \cite{dietl00} and Jungwirth \textit{et al.}\
\cite{PhysRevB.76.125206} indeed predict, within mean-field theory,
that $T_C$ increases monotonically with substitutional Mn concentration and with hole
concentration.

Even after more than a decade, the question of where the holes are
residing is still being debated
\cite{PhysRevLett.97.087208,PhysRevB.73.024411,PhysRevB.76.125206,
PhysRevLett.103.137201,PhysRevLett.105.227201,Sawichi2010,Ohya2011,
Dobrowolska2012,Samarth2012}. One reason for this is that
some experimental probes might be more susceptible to DBH-derived
properties (for example, local probes at Mn sites) while others see more of the
valence-band physics (e.g., transport). Another reason is that
the preparation of heavily Mn-doped GaAs
is a very delicate procedure, which may result in large variations
of sample composition and quality \cite{Nemec2013,PhysRevB.72.125207}.
Furthermore, Mn in GaAs acts as an acceptor,
resulting in the difficulty to
independently control the local-moment and hole concentrations. On the
other hand, interstitial Mn ions (Mn$_\mathrm{I}$) are double donors,
which, intentionally or unintentionally,
compensate the hole doping. Independent precise control of both moment density
and hole concentration is required to resolve the ``band
battle''~\cite{Samarth2012}.

Four different approaches have been applied to adjust the hole
concentration while keeping the Mn concentration constant. Myers
\textit{et al.}\ \cite{PhysRevB.74.155203} utilized the As-antisite effect to
compensate holes in the low-doping regime of 1--2\% Mn. During
growth, they intentionally did not rotate the substrate to obtain
a variation of the As to Ga flux ratio, and thereby of the hole
concentration, across a single wafer. The second approach is to
compensate holes in (Ga,Mn)As by exposing the samples to a
hydrogen plasma \cite{PhysRevLett.92.227202}, followed by
post-annealing
\cite{thevenard2005tuning,khazen2008ferromagnetic,PhysRevB.75.195218}.
In this way, the magnetic and electronic properties of (Ga,Mn)As can be changed
qualitatively, though without fine control. On the other hand, the
third approach, electrical gating of a (Ga,Mn)As metal-insulator-semiconductor
structure, is expected to allow for a fine control over the
hole concentration \cite{ohno_spinFET}, but the required
large gate voltage limits this method to
samples with relatively low $T_C$ \cite{Sawichi2010}. Finally,
it is possible to tune the hole concentration in (Ga,Mn)As by ion irradiation
\cite{PhysRevB.81.045205,PhysRevB.81.245203,0022-3727-44-4-045001}, which is a
standard industry method and allows for a fine tunability and
reproducibility. This method is widely used for conventional
III-V semiconductors to render a conducting layer highly resistive
through the creation of carrier-trapping centers
\cite{Pearton1990313, deenapanray:9123}. It has been demonstrated
to be applicable to (Ga,Mn)As
\cite{PhysRevB.81.045205,PhysRevB.81.245203,0022-3727-44-4-045001}
and (Ga,Mn)P \cite{winkler:012103}.

In this contribution, we aim to shed light on the debate over the
impurity vs.\ valence-band picture for (Ga,Mn)As-based DFS with high Mn
concentration. Our approach is to examine
the magnetic and transport properties while shifting the Fermi level to
higher energies by compensating the free carriers. We demonstrate the
possibility of fine-tuning the magnetism of highly
conducting (Ga,Mn)As films by ion irradiation. With
increasing the displacement per atom (DPA), the (Ga,Mn)As
films become gradually more resistive.
For a broad range of (Ga,Mn)As and also (Ga,Mn)(As,P) samples, we find
universally that $T_C$ \emph{distinctly and monotonically} decreases with
increasing DPA, which we interpret in terms of a decreasing
free-hole concentration. Our samples are clearly distinct from those studied in
Ref.\ \onlinecite{PhysRevB.81.245203}, where the magnetization curves
become more concave and non-Brillouin-function-like with increasing DPA, while
$T_C$ changes relatively little. Even in the strongly insulating
regime, we observe values of $T_C$ in the range 5--10\,K, unlike in
Ref.\ \onlinecite{PhysRevLett.99.227205}, where $T_C$ suddenly drops to 0\,K
from around 10--20\,K.
Taken together with the other observables, e.g., the coercive field and
the shape of the magnetization curves, our observations
are naturally explained in terms of the valence-band picture but not of
the impurity-band picture. This provides support for the former for
ferromagnetic (Ga,Mn)As with high Mn concentrations.

%In our opinion, the impurity band picture describes a \blue{DOS}
%with a pronounced dip between the impurity-derived states and the valence band
%even for samples with high Mn concentrations, while the valence band model
%describes a merged DOS \cite{PhysRevLett.90.029701}.

\section{Experiment}

\begin{table*}
\caption{\label{tab:sample} Sample definitions and related
parameters. The ending in Sample ID: ``ann'' means that the sample was annealed
at low temperatures for long time, while ``ag'' means that the samples was as
grown.}
\begin{ruledtabular}
\begin{tabular}{ccccc}
  Sample ID  & Stoichiometry & Thickness & Nominal Mn concentration & Initial
  $T_C$ \\ \hline
 Mn6ann & (Ga,Mn)As & 25\,nm & 0.06 & 129\,K \\
 Mn6ag & (Ga,Mn)As & 25\,nm & 0.06 & 60\,K \\
 Mn10ann & (Ga,Mn)As & 20\,nm & 0.10 & 150\,K \\
 Mn6P6ann & (Ga,Mn)As$_{0.94}$P$_{0.06}$ & 25\,nm & 0.06 & 125\,K \\
 Mn6P9ann & (Ga,Mn)As$_{0.91}$P$_{0.09}$ & 7\,nm & 0.06 & 80\,K \\
\end{tabular}
\end{ruledtabular}
\end{table*}

\subsection{Sample preparation}

(Ga,Mn)As films were prepared at the University of Nottingham (UoN), and
the Institute of Semiconductors (IOS), Chinese Academy of
Sciences. All samples were grown by low-temperature molecular beam
epitaxy (MBE) on semi-insulating GaAs(001). The details of the growth
of the specific samples is described in the following.

Samples Mn6ann and Mn6ag were grown on low-temperature GaAs
buffer layers with a nominal Mn concentration of $x = 0.06$ and with
a thickness of 25\,nm using a Veeco Mod Gen III MBE system at UoN
\cite{WangM2008}. The Mn concentration was determined from the
Mn/Ga flux ratio. The material quality was found to strongly
depend on the growth temperature and on the As/Ga flux ratio.
The growth temperature of 230\,$^\circ$C is chosen to be the highest
possible while maintaining 2D growth, as monitored by reflection high-energy
electron diffraction. The films were grown under relatively low As flux to
minimize the concentration of As antisites (As$_\mathrm{Ga}$).
The use of As$_{2}$ as a source instead of As$_{4}$ also results
in lower As$_\mathrm{Ga}$ concentrations \cite{campion2003high}. Sample Mn6ann
was annealed in air at 190\,$^\circ$C for 48 hours to remove compensating
Mn$_\mathrm{I}$. Sample Mn6ag was kept in the as-grown state.

The heavily Mn-doped (Ga,Mn)As film, sample Mn10ann, was grown at
200\,$^\circ$C with a nominal Mn concentration of $x = 0.10$ and a
thickness of 20\,nm at IOS \cite{ChenL2009}. During growth, the
As/Ga beam equivalent pressure ratio was set to 8. After growth,
the sample was annealed at 200\,$^\circ$C for 2 hours in air.

P-alloyed (Ga,Mn)(As,P) samples were grown at UoN. A 50\,nm
low-temperature GaAs$_{1-y}$P$_{y}$ buffer layer was grown right
before a 25\,nm Ga$_{0.94}$Mn$_{0.06}$As$_{1-y}$P$_{y}$ layer,
with nominal concentrations $y = 0.06$ and $y = 0.09$ for samples Mn6P6ann and
Mn6P9ann, respectively. The buffer and Mn-containing layers
were all grown at a low temperature of 230\,$^\circ$C. Both were
annealed in air at 190\,$^\circ$C for 48 hours.

The sample definitions and the most important parameters are listed in
Tab.\ \ref{tab:sample}.

\subsection{Ion irradiation}

All (Ga,Mn)As wafers were cut into pieces of dimensions of about
5$\times$5\,mm$^{2}$ for ion irradiation. The He-ion energy was chosen
as 4\,keV (a few as 650\,keV). The details of the irradiation conditions
are given in table \ref{tab:sample_details} in App.\ \ref{app.irr}. During ion
irradiation, the samples were tilted by
7$^{\circ}$ to avoid channeling. The samples were
pasted onto Si wafers, which were kept at around room temperature
without cooling or heating. The resulting defects were found to be
distributed roughly uniformly through the whole (Ga,Mn)As layer by simulation
using the SRIM (Stopping and Range of Ions in Matter) code \cite{srim}.
In the following, we refer to the displacement per atom (DPA), which is
the number of times that an atom in the target is displaced during
irradiation. The DPA is a function of
projectile type (ion species, neutron, or light energetic particles), energy,
fluence, as well as material properties. In the literature, the calculated DPA is often used as a measure of the irradiation
effect in materials. As an example, for GaAs, 14 He ions of 4\,keV
will produce a similar DPA as one Ne ion of 20\,keV at the same depth.
The DPA is a better representation of the effect of irradiation on
materials properties than ion fluence (dosage). It allows for a comparison
between our results and data reported in the literature, in which often
different ion species and energies are used.

By Rutherford backscattering spectroscopy in channeling geometry, which is
sensitive to point defects such as Mn interstitials
\cite{PhysRevB.65.201303,pereira2012stability,PhysRevB.89.115323},
we do not observe measurable increase of Mn interstitials even for the
largest DPA applied in our work (see App.\ \ref{app.c}). In the work
of Winkler \textit{et al.}\ \cite{winkler:012103}, the structural
integrity of (Ga,Mn)P after Ar-ion irradiation was confirmed by several
techniques.
Using ion channeling, the sheet
concentration of Mn$_\mathrm{Ga}$ in their samples
was found to remain constant.
High-resolution transmission electron microscopy and atomic force
microscopy similarly showed no qualitative changes with ion
irradiation. Therefore, we can conclude that the effect of He-ion irradiation
is mainly to introduce deep traps and thereby compensate the
holes. The reduction of hole concentration upon ion
irradiation linearly depends on the ion fluence, i.e., the DPA
\cite{PhysRevB.81.045205,winkler:012103}.

\subsection{Measurements}

Magnetic properties were measured with a superconducting quantum
interference device (Quantum Design, SQUID-MPMS or SQUID-VSM)
magnetometer. To determine $T_C$, we have measured the
temperature-dependent magnetization at a small field of 10 or 20\,Oe
after cooling down in field. Magnetotransport properties were
measured with a magnetic field applied perpendicularly to the film
plane in van der Pauw geometry using a commercial Lakeshore Hall
System. Fields up to 80\,kOe were applied over a wide temperature
range from 2.5\,K to 300\,K.

\section{Results and discussion}

\begin{figure*} \center
\includegraphics[scale=0.65]{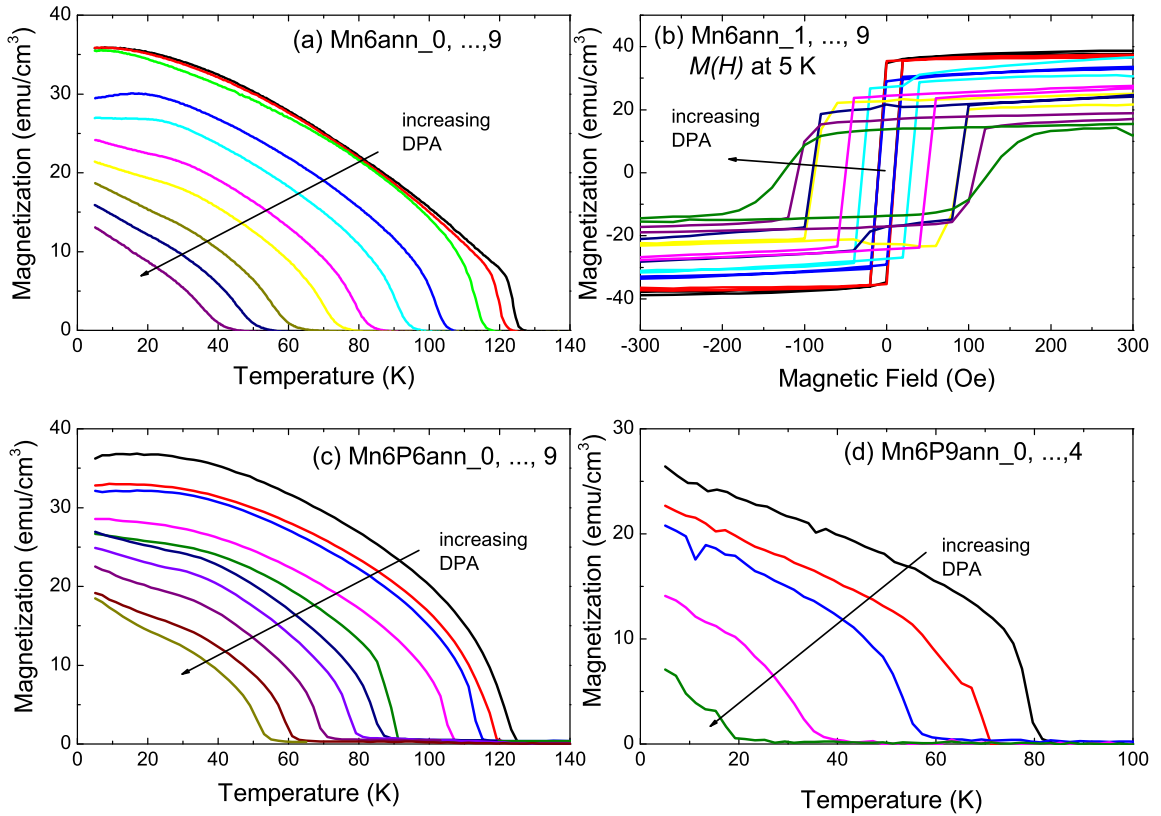}
\caption{Magnetic properties after introducing hole compensation
by ion irradiation. The ion fluence was increased in linear steps.
The hole concentration is expected to decrease linearly depending on DPA
\cite{PhysRevB.81.045205,winkler:012103}.
%For all samples, the magnetization is changed abruptly, revealing
%a distinct change in $T_C$.
Panels (a), (c) and (d) show the temperature dependent magnetization
for different samples, while panel (b) shows the magnetic hysteresis for
sample Mn6ann for various ion fluences. The temperature-dependent magnetization
is measured at a
small field of 20\,Oe after cooling in field. One observes an
increase in the coercive field $H_C$ when $T_C$ and the remanent magnetization
decrease. The arrows indicate the increase of DPA from 0 to
$2.88 \times 10^{-3}$. The corresponding DPA values are shown in
Fig.\ \ref{fig_all_TcHc}. In each panel, the black line is the
result for the non-irradiated sample.}\label{fig1_mag}
\end{figure*}

\begin{figure} \center
\includegraphics[scale=0.3]{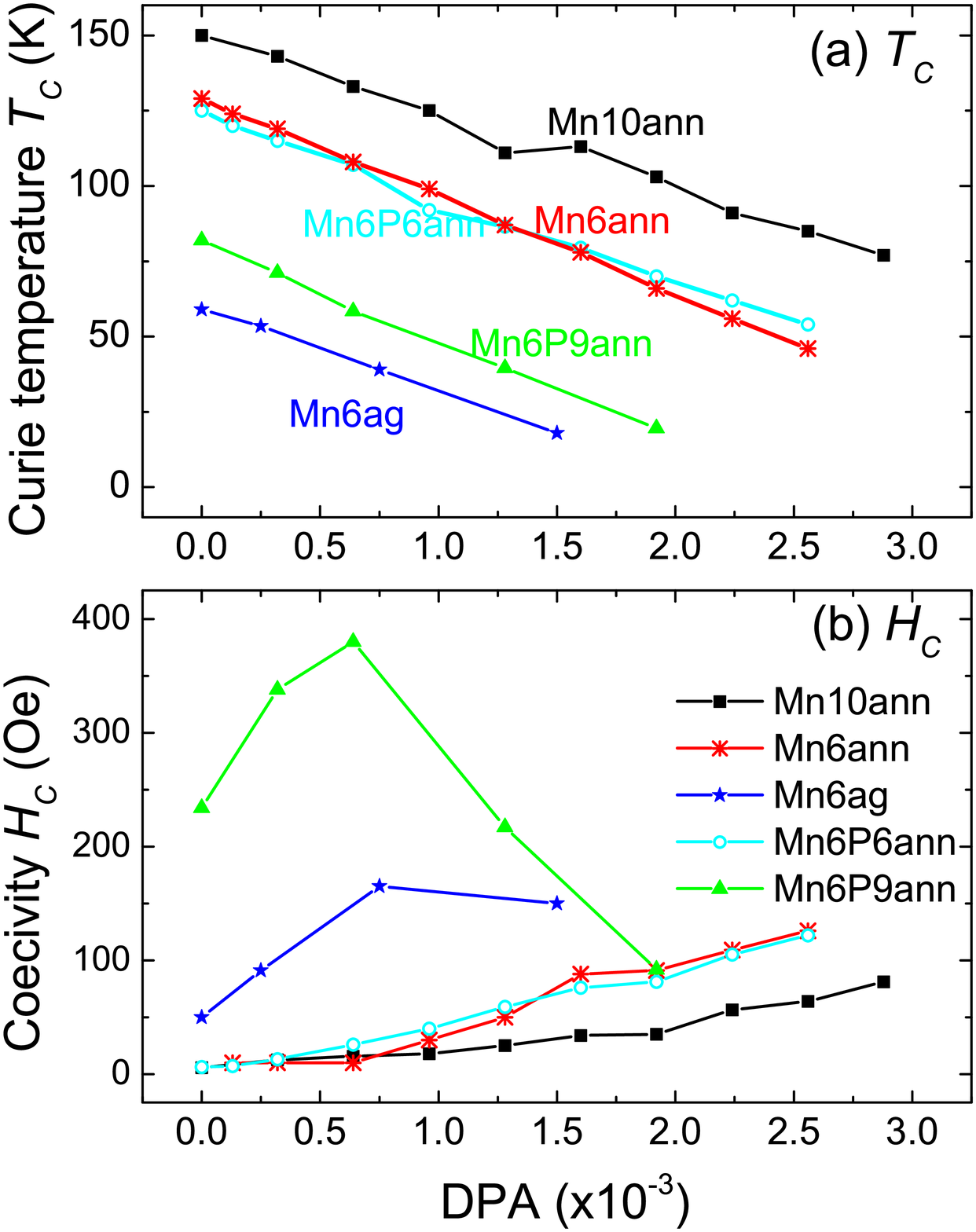}
\caption{(a) Curie temperature $T_C$ vs.\ DPA (ion fluence).
All samples show a monotonic decrease in $T_C$. (b) Coercive field
$H_C$ vs.\ DPA (ion fluence). For the
samples with higher $T_C$, i.e., less compensation (samples Mn6ann,
Mn10ann, Mn6P6ann), $H_C$ shows a monotonic increase with decreasing
hole concentration.}\label{fig_all_TcHc}
% which is in agreement with the valence-band theory
% prediction \cite{PhysRevB.63.054418}.
\end{figure}

Figure \ref{fig1_mag} shows representatively the
temperature-dependent magnetization and the magnetization vs.\
field at 5\,K for three samples listed in Tab.\ \ref{tab:sample}
before and after irradiation at various ion fluences. All
measurements were carried out with the field along the magnetic
easy axis, which is GaAs$[1\bar{1}0]$, except for sample Mn6P9ann,
where it is GaAs[001] (out-of-plane). Figure \ref{fig_all_TcHc}
shows the Curie temperature $T_C$ and the coercive field $H_C$ as
functions of DPA for all samples. Several features can be
observed, which will be discussed in detail below:
\begin{itemize}
\item $T_C$ monotonically decreases with increasing compensation of holes
by ion irradiation over a broad range of DPA.
\item The shape of magnetization vs.\ temperature remains convex from above.
\item There is no significant difference between the behavior of
samples Mn6ann and Mn6P6ann even though the sample Mn6P6ann contains 6\%
P.
\end{itemize}

\subsection{$T_C$ vs.\ compensation}

Let us first discuss the plausible picture regarding the hole
compensation in our samples. We define $p$ as the
absolute hole concentration and $f$ as the ratio
between the hole and Mn concentrations, $f = 0, 0.5, 1$ meaning
full, half, and zero compensation, respectively.
The samples Mn6ann, Mn10ann, and Mn6P6ann show rather high $T_C$ and are only
weakly compensated by Mn$_\mathrm{I}$ or other defects before irradiation.
For samples Mn6ann and Mn6P6ann, $p$
has been carefully measured through the Hall effect, showing that
$f$ is definitely
above 0.5, see Ref.\ \onlinecite{wang2012studies}. Moreover, the saturation
magnetization for sample Mn6ann (non-irradiated) is around 3.32 $\mu_B$/Mn. The
value is nearly the same as that for state-of-the-art high-quality (Ga,Mn)As
\cite{Jungwirth2006lowtemperature}. According to Ref.\
\onlinecite{Jungwirth2006lowtemperature}, the hole concentration is nearly the
same as the effective Mn concentration, $f \approx 1$. The impurity-band
picture naturally predicts a strong suppression of carrier-mediated
ferromagnetism close to $f = 1$, since then the Fermi energy would
fall into a band gap, a mobility gap, or at least an energy range of
reduced DOS between the valence band and the purported impurity band \cite{Dietl_10y}.
Within the impurity-band framework, we should thus see an increase in
$T_C$ as the compensation approaches $f\approx 0.5$.
%
%If we assume p$\leq$0.5, on the other hand we should see the
%temperature-dependent magnetization is changed from a convex shape to a
%concave one \cite{PhysRevB.69.045202}. However, the concave shape of the
%magnetization curve is also an effect of disorder regardless of
%the model and the approximations used \cite{timm2003disorder}.
%Nevertheless, our experimental observations (see Fig.
%\ref{fig1_mag}) dispute the impurity-band picture independent of
%the compensation degree in the initial samples, while it can be
%
Within the mean-field virtual crystal approximation \cite{PhysRevB.72.165204},
$T_C$ is proportional to the Mn$_\mathrm{Ga}$ concentration, to
$p^{1/3}$, and to the effective mass $m^{\ast}$. If the hole-hole exchange
interactions are included, $T_C$ can be slightly enhanced by 10-20\%. Therefore,
in the valence-band picture, $T_C$ is expected to
decrease with decreasing hole concentration, regardless of the initial
compensation. This is indeed what we find: $T_C$ universally decreases
with increasing DPA for all samples. In App.\ \ref{app.b}, we calculate the hole concentration $p$ for sample Mn6ann by following the approach described in Ref. \onlinecite{PhysRevB.81.045205}. Although $T_C$ reveals a linear dependence on $p^{1/3}$, the dependence of $T_C$ on $p$ is insensitive to the exponent 1/3 or 1 in the range of the hole concentration of our samples. Note that the exponent 1/3 only follows for a parabolic dispersion and neglecting disorder \cite{PhysRevB.72.165204}. Our argument does not rely on the specific value of the exponent but on the monotonicity.
%%%This monotonic relationship holds
%%%over a broad range of DPA and the slope appears to be universal.
%
%$T_C$ decreases and the coercive field $H_c$ (see Fig.\ \ref{fig_all_TcHc})
%increases with decreasing hole concentration as predicted by the valence-band
%model \cite{PhysRevB.63.054418}.

\subsection{Phosphorus co-alloying}

Phosphorus co-alloying in (Ga,Mn)As has been employed to tune the magnetic
anisotropy
\cite{lemaitre2008strain,cubukcu2010adjustable,Cubukcu2010,Yahyaoui2013}, the
Fermi energy \cite{Cubukcu20118212}, and the spin stiffness \cite{Shihab2015}.
It is instructive to consider samples Mn6P6ann and Mn6P9ann, which
are co-alloyed with 6\% and 9\% phosphorus, respectively. In the
nonirradiated samples, we do not see significant differences in
magnetization and $T_C$ between samples Mn6P6ann and Mn6ann, which contains
the same nominal concentration of Mn but no P. This
strongly differs from
the work by Stone \textit{et al.}\ \cite{PhysRevLett.101.087203},
who observed a significant reduction of $T_C$ upon P co-alloying.
They also found that ferromagnetic
Ga$_{0.954}$Mn$_{0.046}$As films are changed from metallic to
insulating by around 2.4\% P co-alloying, which is attributed to the fact
that holes located within an impurity band are scattered by alloy
disorder in the anion sublattice \cite{PhysRevLett.101.087203}.
However, our P-alloyed sample Mn6P6ann is metallic even though the P
concentration is 6\%. Moreover, comparing samples
Mn6ann with Mn6P6ann, our results reveal similar dependence of magnetization and $T_C$ on the hole compensation. On the other hand, sample
Mn6P9ann exhibits a remarkable reduction in $T_C$ compared to samples
Mn6ann and Mn6P6ann, likely due to its larger concentration of P.
In the work by Cubukcu \textit{et al.}\ \cite{Cubukcu20118212},
the energy level of Mn moves deeper into the gap and
the transport character changes from metallic to insulating
with increasing P concentration. The Curie temperature drops sharply
from 130\,K to 45\,K when the P concentration is increased from 0 to around
19\%. The extreme representatives
are (Ga,Mn)P and (In,Mn)P, in which the Mn energy levels are located 400\,meV and
220\,meV above the valence-band
edge, respectively, compared to 110\,meV for (Ga,Mn)As. Both materials
exhibit hopping conduction
\cite{PhysRevLett.95.207204,PhysRevB.89.121301}.

A theoretical work by R. Bouzerar \textit{et al.}\
\cite{PhysRevB.82.035207} explains the large reduction of $T_C$ for P
co-alloyed (Ga,Mn)As reported in Ref.\ \onlinecite{PhysRevLett.101.087203}.
They employ the local spin-density approximation to obtain Mn-Mn exchange
interactions. These interactions are used in a Heisenberg model with random Mn
positions, which is treated by the selfconsistent random-phase approximation
\cite{0295-5075-69-5-812} and complementary Monte Carlo simulations.
Compensating defects---P antisites as well as Mn and P interstitials---lead
to the reduction of the hole concentration and of $T_C$ compared to
(Ga,Mn)As not containing P. If compensating defects can be avoided,
the change in $T_C$ is expected to be negligible for low P concentration
\cite{rushforth:073908,Cubukcu20118212}, which is indeed in good agreement
with our MBE-grown samples.

\subsection{Magnetization and coercivity vs.\ compensation}
\label{sub_mag}

The reduction of the remanent magnetization $M$ at low
temperatures, shown in Fig.\ \ref{fig1_mag}, is still
puzzling. It cannot be explained by a loss of substitutional Mn
ions. Note that we applied He-ion irradiation with very low
fluences ($10^{12}$--$10^{13}\,\mathrm{cm}^{-2}$). The maximal DPA is only
0.2--0.3\% according to SRIM simulations \cite{srim}. Most of the substitutional
Mn ions are expected to remain at their original sites, as we
do not see any change in the channeling analysis along GaAs[001] or [110],
see App.\ \ref{app.c}. Therefore, within the valence-band framework the
magnetization at zero temperature for the compensated samples is
expected to stay the same as for the initial sample. Yet we see a
significant reduction in both the remanent and the low-temperature
saturation magnetization. At the largest fluence applied to sample
Mn6ann, see Fig.\ \ref{fig1_mag}(b), the magnetization is reduced by more
than 50\%. The reduction of magnetization has been also observed in
hydrogenated \mbox{(Ga,Mn)As} with reduced hole concentration \cite{khazen2008ferromagnetic} and can be reversed by annealing
\cite{PhysRevB.75.195218}. This phenomenon can be
understood by considering the
inhomogeneity of (Ga,Mn)As as suggested by Dietl
\cite{dietl2008interplay} and by Sawicki \textit{et al.}\
\cite{Sawichi2010}. Carrier-mediated ferromagnetism is strongly
influenced by the vicinity of the insulator-to-metal transition
(IMT). Particularly, the spatial fluctuations in the
local DOS can destroy the spatial homogeneity of the
ferromagnetic order, resulting in a nanoscale phase separation. We
consider a similar scenario: when defects are introduced,
regions with an initially small hole concentration are depleted first
and become paramagnetic or superparamagnetic.
Therefore, a nanoscale phase separation regarding both
the effective Mn concentration and the hole concentration can
lead to the gradual appearance of superparamagnetic and paramagnetic
regions, consequently reducing the total magnetization. Indeed, we have
observed a superparamagnetic component in sample Mn6ag with the lowest
initial $T_C$ after
irradiation up to large He fluences by measuring the
zero-field-cooled/field-cooled temperature-dependent
magnetization (see App.\ \ref{app.a}).

%From the values shown in
%Fig.\ \ref{fig1_mag}(b), the magnetization at 5\,K is reduced step
%by step upon introducing compensation.

Another reason for a decrease in the magnetization is the
occurrence of antiferromagnetically coupled nearest-neighbor
Mn$_\mathrm{Ga}$ pairs when the sample is heavily compensated
\cite{PhysRevB.69.115208,PhysRevB.72.165204}. The antiferromagnetically coupled
pairs are invisible in the magnetization even at larger fields since the
short-range antiferromagnetic interaction is large compared to the Zeeman
energy.

The reduction of the remanent and apparent saturation magnetization with
hole compensation naturally leads to an increase in the coercive
field $H_C$. A consistent increase in $H_C$ with DPA for samples
Mn6ann, Mn10ann, and Mn6P6ann is indeed seen in Fig.\
\ref{fig_all_TcHc}. The increase in $H_C$ agrees with predictions
obtained from a valence-band description
\cite{PhysRevB.63.054418}, lending additional support to this
picture. Note that $H_C$ starts to decrease for samples Mn6P9ann and
Mn6ag when the samples are subjected to much larger DPA. It is due
to the fact that $T_C$ for those samples are already very low and
only slightly above 5\,K, the measurement temperature for $H_C$
shown in Fig.\ \ref{fig_all_TcHc}. On the other hand, the magnetic anisotropy
can also change when the compensation is high
\cite{khazen2008ferromagnetic,PhysRevB.75.195218,0022-3727-44-4-045001}. In
hydrogenation-compensated (Ga,Mn)As, the coercivity for magnetic
field in the in-plane direction also increases \cite{PhysRevB.75.195218}.

\begin{figure} \center
\includegraphics[scale=0.4]{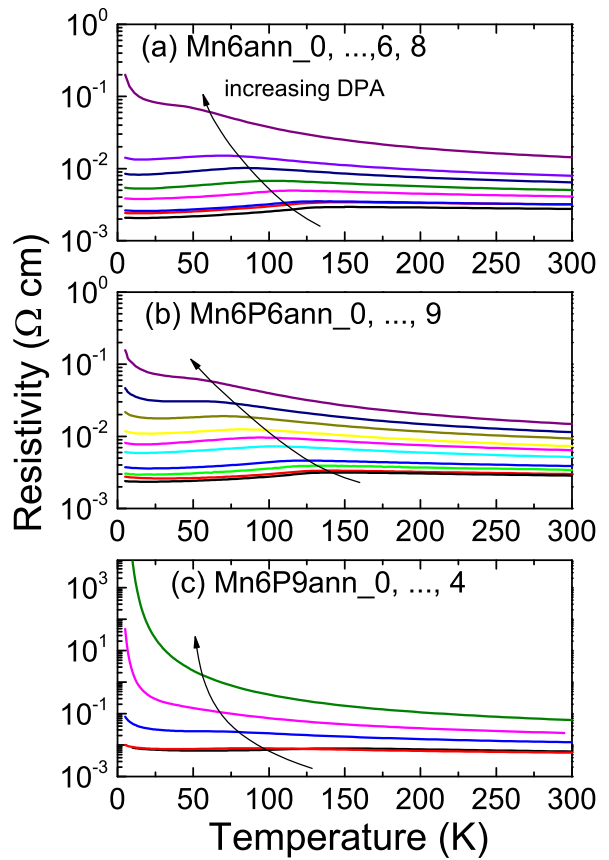}
\caption{Temperature-dependent resistivity in zero field for
three samples upon ion irradiation. For samples Mn6ann (a) and
Mn6P6ann (b), the resistivity continuously increases with decreasing hole
concentration but still shows a broadened cusp around $T_C$, while
sample Mn6P9ann (c) is closer to the insulating regime.
%Note the \blue{very different relative change} of the resistivity at low
%temperatures \blue{with compensation}:
%for sample Mn6P6ann it \blue{changes} by less than two orders of magnitude,
%\blue{whereas for sample Mn6P9ann it changes by} more than
%five orders of magnitude.
%These samples are the same as those shown in Fig.\ \ref{fig_all_TcHc}.
The arrows indicate the increase of the DPA from 0 to
$2.88 \times 10^{-3}$. The corresponding DPA values are shown in
Fig.\ \ref{fig_all_TcHc}.}\label{Fig_RvsT}
\end{figure}

\subsection{Resistivity vs.\ compensation}

Corresponding effects are observed in the
temperature-dependent resistivity, as shown in Fig.\ \ref{Fig_RvsT}.
We present data for three samples: Mn6ann, Mn6P6ann and Mn6P9ann.
Mn6P9ann is closer to the insulating regime due to the high
concentration of P. With increasing compensation, the resistivity for samples
Mn6ann and Mn6P6ann continuously increases and develops an upturn at low
temperatures, but both samples still show a broadened cusp around the Curie
temperature. For sample Mn6P9ann, the cusp is present for low ion fluence
but is overwhelmed by a huge low-temperature increase of the resistivity for
moderate fluences.
Note the very different relative change of the resistivity value at low temperatures:
for samples Mn6ann and Mn6P6ann it changes by less than two orders of magnitude,
whereas for sample Mn6P9ann it changes by more than five orders of magnitude.
We have also measured the Hall
resistance at 300\,K (above $T_C$) for our samples, examples
are shown in App.\ \ref{app.b}. After irradiation, all samples still show
p-type conductivity. The Hall coefficient at 300\,K increases gradually
with increasing DPA, see Fig.\ \ref{fig:Mn377_Hall300K} in App.\
\ref{app.b}. The results indicate a decrease of the hole
concentration, but do not give an
accurate measurement due to the paramagnetic component in the Hall
effect \cite{wang2012studies}. However, Mayer \textit{et al.}\
\cite{PhysRevB.81.045205} have used
electrochemical capacitance voltage profiling to measure the
hole concentration and have confirmed a linear correlation between
hole concentration and DPA. We conclude that the moderate increase of
resistivity in sample Mn6P6ann is mainly due to the reduction of the free
hole concentration, while only at low temperatures an upturn due
to carrier localization develops. On the other hand, the large resistivity
increase for sample Mn6P9ann with a low initial $T_C$
is dominated by localization.

%In Table II, we list the typical observable parameters in magnetic
%properties for ferromagnetic (Ga,Mn)As or (Ga,Mn)AsP upon hole
%compensation. The expected results from the two models are also
%shown for comparison. Our experimental results strongly support
%the valence-band model: (1) $T_C$ monotonically decreases with decreasing hole
%concentration; (2) in the low compensation regime $H_c$ decreases with
%increasing Mn concentration and hole concentration; (3) the temperature
%dependent magnetization follows a mean-field like behavior and (4) there is
%only
%moderate modification by co-alloying with phosphorus at low concentration.

\subsection{Onset of ferromagnetism}

\begin{figure*} \center
\includegraphics[scale=0.6]{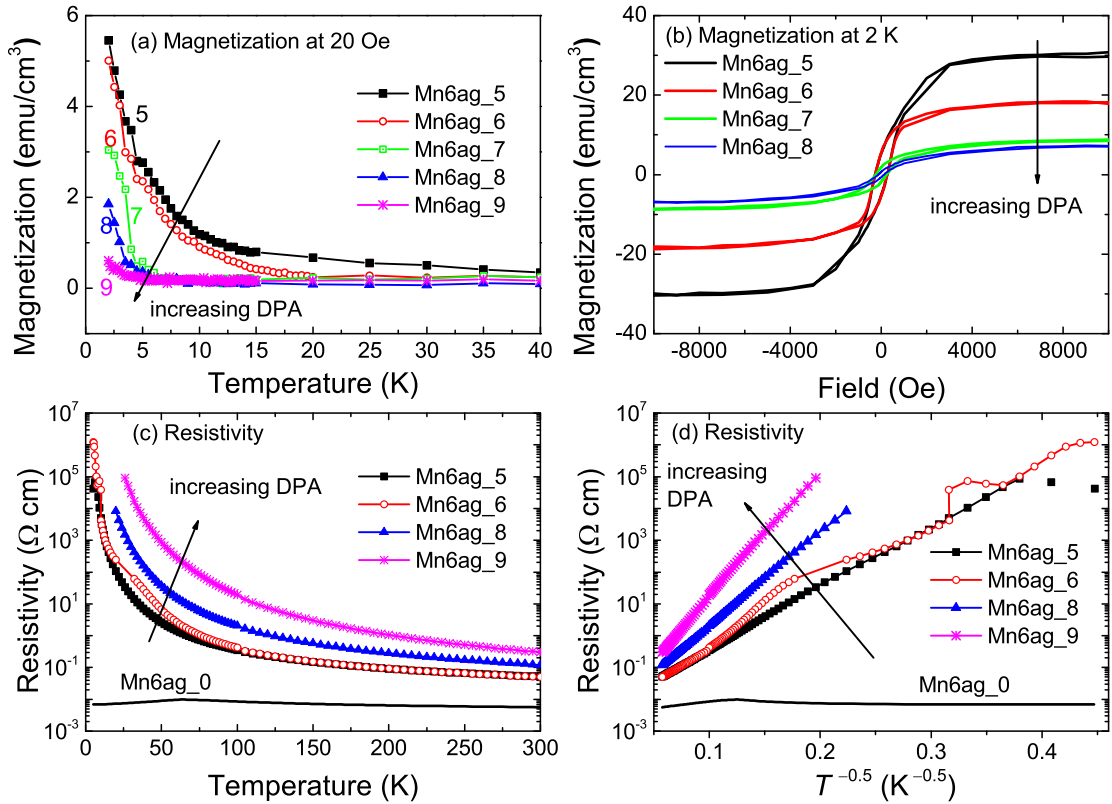}
\caption{Change of magnetic and transport properties of sample Mn6ag with
compensation, close to the destruction of ferromagnetism.
(a) Temperature dependent magnetization in a field of 20 Oe, (b) Magnetization vs.\
field at 2\,K.
(c) and (d) temperature-dependent longitudinal resistance in zero field.
A virgin, nonirradiated sample (Mn6ag\_0) is shown for comparison. Mn6ag\_5,
\ldots, Mn6ag\_9 label the samples with different DPA.}\label{fig_onset}
\end{figure*}

Within the Zener \emph{p-d} exchange scenario, the Curie
temperature should continuously decrease with decreasing hole
concentration even into the insulating regime, since the electronic properties
at the length scale of the typical Mn-Mn separation change
continuously with hole concentration through the IMT
\cite{Dietl_10y,Sawichi2010,dietl00}. Actually, Curie temperatures
as low as 0.75\,K to 2.4\,K for p-doped (Zn,Mn)Te
\cite{PhysRevB.63.085201} and from 5\,K to 8\,K for (In,Mn)Sb
\cite{Wojtowicz2003} were measured.

Within the impurity-band picture
\cite{PhysRevLett.99.227205}, the interaction between Mn ions is double-exchange
via the hopping of holes between two Mn ions. Therefore, the free-hole
concentration $p$ has only a weak effect on $T_C$. Instead,
the hopping energy should be a good predictor for ferromagnetism
\cite{PhysRevLett.99.227205}. However, while the experimental data
\cite{PhysRevLett.99.227205} show that ferromagnetic order only occurs within a
certain range of hopping energies, the correlation between $T_C$ and the
hopping energy is weak. The theory framework described in Ref.\
\cite{PhysRevLett.99.227205} predicts a discontinuous disappearance of the
ferromagnetic state. Indeed, the experimental $T_C$ for (Ga,Mn)As
is reported to jump from 10--15\,K to zero.

%Moreover, in Ref.\
%\onlinecite{PhysRevLett.99.227205} the authors indeed claim that the hopping
%energy is a good predictor for $T_C$, but this claim is not supported by their
%experimental data. There is no direct correlation between the $T_C$ and the
%hopping energy for the ferromagnetic samples.

The Anderson-Mott IMT occurs at a critical carrier concentration
$p_c$. Empirically, the magnitude of $p_c$ is in the range
$p_c^{1/3}a_B = 0.26\pm0.05$ \cite{RevModPhys.86.187}, where $a_B$
is the effective Bohr radius. $a_B$ can be evaluated from the
impurity binding energy $E_I$ by $a_B = \hbar/(2m^*E_I)^{1/2}$ or
$a_B = e^2/8\pi\epsilon_0\epsilon_rE_I$
\cite{PhysRevB.17.2575,RevModPhys.86.187}. For details, see Ref.\
\onlinecite{RevModPhys.86.187}. For GaAs, the Mn single-impurity
binding energy is 110 meV, but for InSb it is as low as several
meV. Therefore, one can expect a much larger $p_c$ for (Ga,Mn)As
than for (In,Mn)Sb. The double role of Mn ions in GaAs as both
local spins and acceptors usually prevents fine tuning of the
carrier concentration near the IMT. This may explain the absence
of the observation of a continuous decay of $T_C$ to 0\,K for
(Ga,Mn)As. We expect that a (Ga,Mn)As sample with high Mn
concentration but with very low hole concentration shows a $T_C$
close to zero. We choose sample Mn6ag, which has a relatively low
initial $T_C$. Five pieces of sample Mn6ag were irradiated with He
ions. The ion fluence was increased in very small step starting
from a DPA of 1.5$\times$10$^{-3}$ shown in Fig.\
\ref{fig_all_TcHc}(a), which leads to the hole concentration
decreasing in small steps. Figure \ref{fig_onset} shows the
magnetic and transport properties of this set of samples. For all
of them, the temperature-dependent remanent magnetization is
\emph{concave} from above, distinct from Fig.\ \ref{fig1_mag}.
This is typical for DFS with Fermi energy deep in the localized
states in the disorder-induced band tail \cite{timm2003disorder}.
For these concave magnetization curves, it is difficult to extract
precise values for $T_C$. Nevertheless, $T_C$ in the range 5--10\,K is
observed. It is more
instructive to consider the magnetization in field measured at
2\,K, shown in Figure \ref{fig_onset}(b). The samples still show
clear ferromagnetic hysteresis and the saturation magnetization
decreases gradually with increasing DPA. This is consistent with
the suppression of ferromagnetism with decreasing hole
concentration. We note that this set of samples all show
insulating behavior, consistent with strong disorder. The
conductivity can be understood as hopping as shown
in Figs.\ \ref{fig_onset}(c) and \ref{fig_onset}(d) by fitting to
a $T^{-0.5}$ dependence \cite{efros1984electronic}.

Our results provide direct experimental evidence that
ferromagnetism can be suppressed with $T_C$ below 10 K by introducing
hole compensation in (Ga,Mn)As samples with large Mn
concentration. This is easy to understand within the valence-band
picture. In sample Mn6ag with $x = 0.06$, the Mn-derived impurity band is merged with
the valence band before we introduce hole compensation. If we
introduce compensation, the Fermi energy shifts up into the tail
of the valence band. With decreasing hole concentration, the Fermi
energy shifts deeper and deeper into this tail. States
at the Fermi energy are thus more and more localized.
Ferromagnetism is mainly determined by the properties of the hole states
on the length scale of the typical Mn-Mn nearest-neighbor distance, which
change smoothly with the localization length and thus with compensation.
Since the Mn concentration is high, the Mn-Mn distance is typically smaller than the
effective Bohr radius of an acceptor-bound state. In this sense,
the Zener \emph{p-d} exchange can still happen, leading to the
formation of Zhang-Rice magnetic polarons \cite{PhysRevB.66.033203,PhysRevLett.88.247202}.
The temperature below which the magnetic polarons overlap and
ferromagnetic order develops depends on the hole concentration.
Consequently, $T_C$ can be decreased to close to zero Kelvin
when the hole concentration is gradually reduced.

\section{Summary and conclusions}

We have presented the magnetic and transport properties of a broad
range of (Ga,Mn)As-based DFS samples. The DFS samples used in this
study are of state-of-the-art quality, as shown by their high $T_C$
and confirmed by X-ray diffraction measurements
\cite{PhysRevB.87.121301,wang2012studies}. We have observed a
monotonic decrease of $T_C$ with decreasing hole
concentration over a large temperature range and the existence of Curie
temperatures below 10 K for heavily compensated samples. These observations
cannot be explained if there is a pronounced dip in the DOS between the
Mn impurity derived state and the GaAs valence band. Our results rather support
the valence-band picture for
high-quality (Ga,Mn)As-based DFS. But how can we then understand the
experimental results
reported in Refs.\ \onlinecite{PhysRevLett.101.087203,
PhysRevB.81.045205,PhysRevB.81.245203,Dobrowolska2012}, which are
not consistent with the valence-band picture? We first note that in
Refs.\ \onlinecite{PhysRevB.81.245203,PhysRevB.81.045205,Dobrowolska2012},
the total Mn concentrations are 5\%, 4.5\%, and 6.8\%, but the
corresponding $T_C$ is only around 60\,K, 80\,K, and 90\,K,
respectively. This means that the starting materials were
already highly compensated by defects. The large influence of a small
concentration of P
reported in Ref.\ \onlinecite{PhysRevLett.101.087203} was
later explained by the presence of additional defects arising with
P co-alloying by ion implantation and pulsed laser annealing
\cite{PhysRevB.82.035207}. Given the fact that the preparation of
ferromagnetic (Ga,Mn)As is a rather delicate procedure, a small
deviation in growth parameters can significantly alter the
properties of the prepared materials \cite{Nemec2013}. A reliable
conclusion can only be obtained by working with highest-quality
(Ga,Mn)As materials \cite{edmonds2012correspondence}.

Finally, we return to the question alluded to in the beginning: Is
there hope to realize DFS with $T_C$ above room
temperature? Within the valence-band picture with \emph{p-d}
exchange, based on the present results, we suggest to search for materials (a) with
smaller lattice constants, which lead to a larger
\emph{p-d} exchange \cite{Dietl_10y} and (b) with
a large concentration of mobile holes. The criterion (a) is usually
positively correlated with a wide
bandgap of the host semiconductor. For Mn in III-V semiconductors,
unfortunately the two criteria cannot be fulfilled
simultaneously, since the \emph{p-d} hybridization increases the
binding energy of Mn acceptors going from arsenides to
phosphides and finally nitrides. In samples with strong \emph{p-d}
hybridization, the holes are more localized. Moreover, it is generally
difficult to achieve p-type doping in wide-bandgap semiconductors, particularly
in GaN and ZnO \cite{Wei2004337,Zunger2003}, even by doping with shallower
acceptors than Mn. Searching for different transition-metal
dopants is less promising, since other transition-metal ions
substituted for the cation tend to have several energy levels of both
donor and acceptor type. It has turned out that up
to now (Ga,Mn)As yields the best compromise and has achieved the
highest $T_C$ \cite{Dietl_10y,bouzerar2010unified,peng2013chemical}.
Nevertheless, an intentional introduction of nanoscale inhomogeneities has been
proposed as a path to achieve high Curie temperatures in diluted systems
\cite{kuroda2007origin,Bouzerar2004APL,chakraborty2012nanoscale}. Moreover,
Mn-doped I-II-V compounds \cite{deng2011li} with decoupled spin and charge
doping might provide an alternative
test bed for producing DFS with high $T_C$. At the present
stage, $T_C$ reaches 220\,K in
(Ba$_{0.7}$K$_{0.3}$)(Zn$_{0.85}$Mn$_{0.15}$)$_{2}$As$_{2}$
\cite{zhaoferromagnetism}.

\acknowledgments

Support by the Ion Beam Center (IBC) at HZDR is gratefully acknowledged. Financial support by the Helmholtz Association (VH-NG-713) is gratefully
acknowledged. C.T. acknowledges financial support by the Deutsche
Forschungsgemeinschaft.

\appendix
\section{He ion irradiation}
\label{app.irr}

The samples listed in table \ref{tab:sample} were irradiated with He
ions. The sample notation and experimental condition are listed in table
\ref{tab:sample_details}.

\begin{table}
\caption{\label{tab:sample_details} Sample notations, energy and fluence of He
ions.}
\begin{ruledtabular}
\begin{tabular}{cccc}
 Samples  & Energy & Fluence (cm$^{-2}$) & Peak DPA \\ \hline
 Mn6ann\_0 & -- & 0 & 0 \\
 Mn6ann\_1 & 4 keV & 2.0$\times$10$^{12}$ & 0.13$\times$10$^{-3}$ \\
 Mn6ann\_2 & 4 keV & 5.0$\times$10$^{12}$ & 0.32$\times$10$^{-3}$ \\
 Mn6ann\_3 & 4 keV & 1.0$\times$10$^{13}$ & 0.64$\times$10$^{-3}$ \\
 Mn6ann\_4 & 4 keV & 1.5$\times$10$^{13}$ & 0.96$\times$10$^{-3}$ \\
 Mn6ann\_5 & 4 keV & 2.0$\times$10$^{13}$ & 1.28$\times$10$^{-3}$ \\
 Mn6ann\_6 & 4 keV & 2.5$\times$10$^{13}$ & 1.60$\times$10$^{-3}$ \\
 Mn6ann\_7 & 4 keV & 3.0$\times$10$^{13}$ & 1.92$\times$10$^{-3}$ \\
 Mn6ann\_8 & 4 keV & 3.5$\times$10$^{13}$ & 2.24$\times$10$^{-3}$ \\
 Mn6ann\_9 & 4 keV & 4.0$\times$10$^{13}$ & 2.56$\times$10$^{-3}$ \\

 \hline
 Mn6ag\_0 & -- & 0 & 0 \\
 Mn6ag\_1 & 650 keV & 1.0$\times$10$^{14}$ & 0.25$\times$10$^{-3}$ \\
 Mn6ag\_2 & 650 keV & 3.0$\times$10$^{14}$ & 0.75$\times$10$^{-3}$ \\
 Mn6ag\_3 & 650 keV & 6.0$\times$10$^{14}$ & 1.50$\times$10$^{-3}$ \\
 Mn6ag\_4 & 650 keV & 10.0$\times$10$^{14}$ & 2.50$\times$10$^{-3}$ \\
 Mn6ag\_5 & 4 keV & 2.0$\times$10$^{13}$ & 1.28$\times$10$^{-3}$ \\
 Mn6ag\_6 & 4 keV & 2.5$\times$10$^{13}$ & 1.60$\times$10$^{-3}$ \\
 Mn6ag\_7 & 4 keV & 3.0$\times$10$^{13}$ & 1.92$\times$10$^{-3}$ \\
 Mn6ag\_8 & 4 keV & 3.5$\times$10$^{13}$ & 2.24$\times$10$^{-3}$ \\
 Mn6ag\_9 & 4 keV & 4.0$\times$10$^{13}$ & 2.56$\times$10$^{-3}$ \\
 \hline

 Mn10ann\_0 & -- & 0 & 0 \\
 Mn10ann\_1 & 4 keV & 5.0$\times$10$^{12}$ & 0.32$\times$10$^{-3}$ \\
 Mn10ann\_2 & 4 keV & 1.0$\times$10$^{13}$ & 0.64$\times$10$^{-3}$ \\
 Mn10ann\_3 & 4 keV & 1.5$\times$10$^{13}$ & 0.96$\times$10$^{-3}$ \\
 Mn10ann\_4 & 4 keV & 2.0$\times$10$^{13}$ & 1.28$\times$10$^{-3}$ \\
 Mn10ann\_5 & 4 keV & 2.5$\times$10$^{13}$ & 1.60$\times$10$^{-3}$ \\
 Mn10ann\_6 & 4 keV & 3.0$\times$10$^{13}$ & 1.92$\times$10$^{-3}$ \\
 Mn10ann\_7 & 4 keV & 3.5$\times$10$^{13}$ & 2.24$\times$10$^{-3}$ \\
 Mn10ann\_8 & 4 keV & 4.0$\times$10$^{13}$ & 2.56$\times$10$^{-3}$ \\
 Mn10ann\_9 & 4 keV & 4.5$\times$10$^{13}$ & 2.88$\times$10$^{-3}$ \\

 \hline
 Mn6P6ann\_0 & -- & 0 & 0 \\
 Mn6P6ann\_1 & 4 keV & 2.0$\times$10$^{12}$ & 0.13$\times$10$^{-3}$ \\
 Mn6P6ann\_2 & 4 keV & 5.0$\times$10$^{12}$ & 0.32$\times$10$^{-3}$ \\
 Mn6P6ann\_3 & 4 keV & 1.0$\times$10$^{13}$ & 0.64$\times$10$^{-3}$ \\
 Mn6P6ann\_4 & 4 keV & 1.5$\times$10$^{13}$ & 0.96$\times$10$^{-3}$ \\
 Mn6P6ann\_5 & 4 keV & 2.0$\times$10$^{13}$ & 1.28$\times$10$^{-3}$ \\
 Mn6P6ann\_6 & 4 keV & 2.5$\times$10$^{13}$ & 1.60$\times$10$^{-3}$ \\
 Mn6P6ann\_7 & 4 keV & 3.0$\times$10$^{13}$ & 1.92$\times$10$^{-3}$ \\
 Mn6P6ann\_8 & 4 keV & 3.5$\times$10$^{13}$ & 2.24$\times$10$^{-3}$ \\
 Mn6P6ann\_9 & 4 keV & 4.0$\times$10$^{13}$ & 2.56$\times$10$^{-3}$ \\
 \hline
 Mn6P9ann\_0 & -- & 0 & 0 \\
 Mn6P9ann\_1 & 4 keV & 5.0$\times$10$^{12}$ & 0.32$\times$10$^{-3}$ \\
 Mn6P9ann\_2 & 4 keV & 1.0$\times$10$^{13}$ & 0.64$\times$10$^{-3}$ \\
 Mn6P9ann\_3 & 4 keV & 2.0$\times$10$^{13}$ & 1.28$\times$10$^{-3}$ \\
 Mn6P9ann\_4 & 4 keV & 3.0$\times$10$^{13}$ & 2.56$\times$10$^{-3}$ \\

\end{tabular}
\end{ruledtabular}
\end{table}

\section{Rutherford backscattering spectrometry/channeling}
\label{app.c}

\begin{figure} \center
\includegraphics[scale=0.3]{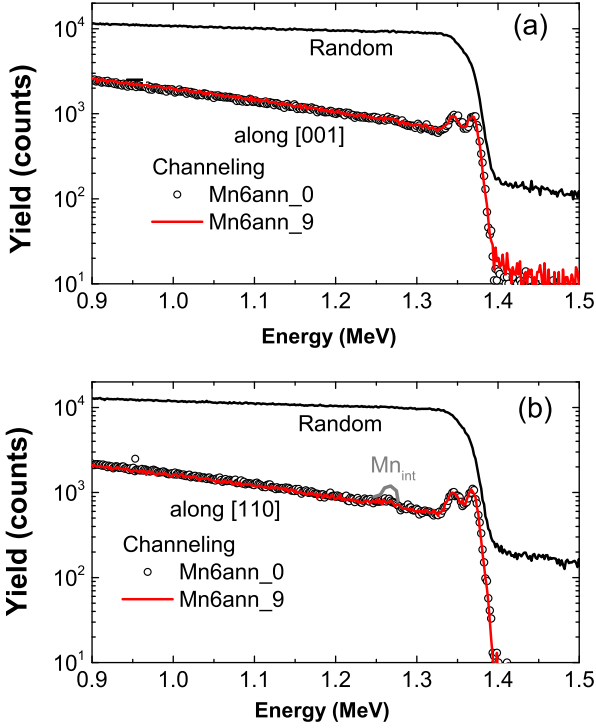}
\caption{RBS/Channeling results for sample Mn6ann, comparing virgin (Mn6ann$\_0$) and irradiated (Mn6ann$\_9$) samples. The channeling spectra do not reveal any
detectable increase in yield along both [001] (a) and [110] (b). The grey curve schematically drawn in (b) indicates the expected spectrum if there is a lot of Mn interstitials after irradiation.}\label{fig:RBS}
\end{figure}

Rutherford backscattering spectrometry/channeling (RBS) is sensitive to
the crystalline disorder and is able to quantitatively measure the
fraction of substitutional and interstitial Mn ions
\cite{PhysRevB.65.201303,PhysRevB.89.115323}. Since the interstitial Mn ions can be invisible along GaAs[001] \cite{PhysRevB.65.201303,PhysRevB.89.115323}, we measure the channeling spectra along both GaAs[001] and [110] axis. The RBS measurements
were performed with a collimated 1.7\,MeV He$^+$ beam at a
backscattering angle of 170$^{\circ}$. The sample was mounted on a
three-axis goniometer with a precision of 0.01$^{\circ}$. The channeling
spectra were measured by aligning the samples to make the impinging
He$^+$ beam parallel to the GaAs[001] or [110] axis. Figure
\ref{fig:RBS} shows a measurement for samples Mn6ann before and
after ion irradiation (up to the largest He ion fluence). Within the detection limit, there is no
signature of an increase in the number of interstitial Mn ions. In Fig. \ref{fig:RBS}(b), the expected channeling spectrum \cite{PhysRevB.89.115323} in the presence of interstitial Mn ions is schematically drawn on top of the experiment spectrum.

\begin{figure} \center
\includegraphics[scale=0.3]{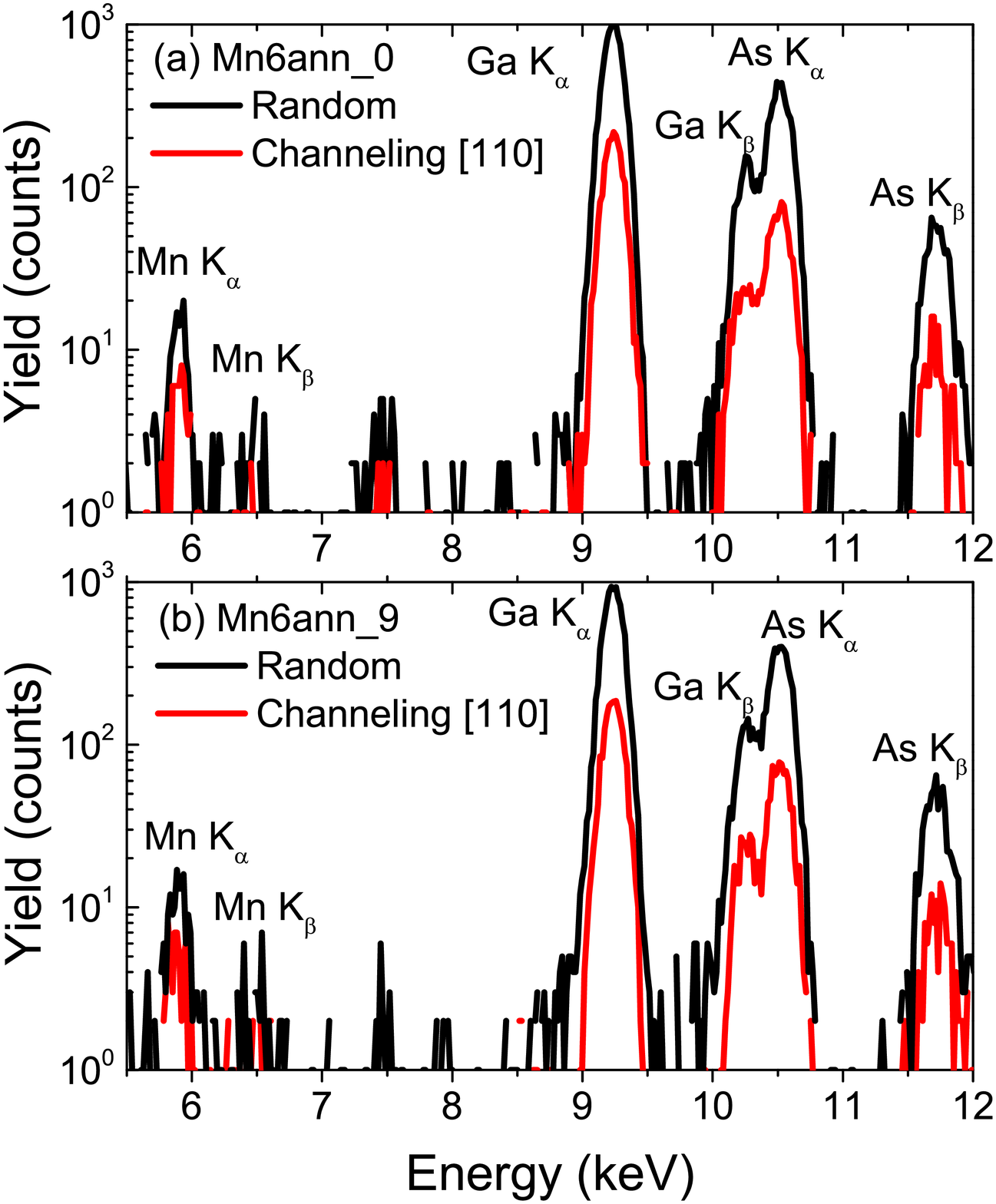}
\caption{PIXE/Channeling results for sample Mn6ann along GaAs[110]: (a) Mn6ann$\_0$ before irradiation and (b) Mn6ann$\_9$ after irradiation with largest fluence. In both samples, the channeling effect is clearly revealed.}\label{fig:PIXE}
\end{figure}

Moreover, the particle-induced X-ray emission (PIXE) was recorded simultaneously during RBS measurements. The advantage is that the Mn \emph{K}-lines are well seperated from the Ga and As \emph{K}-lines. However, the emission is relatively weak since we use He ions instead of proton \cite{PhysRevB.65.201303}. Figure \ref{fig:PIXE} shows the PIXE/channeling spectra along GaAs[110] for both samples Mn6ann$\_0$ and Mn6ann$\_9$. Although the signal from Mn \emph{K}-lines is weak and noisy, the channeling effect is clearly revealed: the emission is much dropped when the beam is along GaAs[110] (the red spectra in the figure). One can roughly estimate the nonrandom fraction ($f_{nr}$) of Mn using $f_{nr}=(1-{\chi}_{Mn}/(1-{\chi}_{Ga})$. Here ${\chi}_{Mn}$ (${\chi}_{Ga}$) is the ratio between the integral intensity of the Mn $K_{\alpha}$ (Ga $K_{\alpha}$) at the channeling geometry and at the random geometry. For both samples, $f_{nr}$ is estimated to be around 75\%. This number can be considered as a measure of the substitutional fraction of Mn. However, $f_{nr}$ can be smaller than the Mn substitutional fraction due to the flux peaking effect \cite{PhysRevB.65.201303,PhysRevB.89.115323}. Nevertheless, $f_{nr}$ along GaAs[110] remains the same after He ion irradiation up to the largest fluence.

\section{Superparamagnetism in highly compensated (Ga,Mn)As}
\label{app.a}

As noted in Sec.\ \ref{sub_mag}, carrier-mediated ferromagnetism is
strongly influenced by the vicinity of the IMT.
In particular, spatial fluctuations in the effective Mn and hole
concentrations are expected to lead to the formation of paramagnetic or
superparamagnetic regions, as the hole concentration is reduced. We
have indeed observed signatures for paramagnetic and
superparamagnetic components in sample Mn6ag after irradiation.

\begin{figure} \center
\includegraphics[scale=0.3]{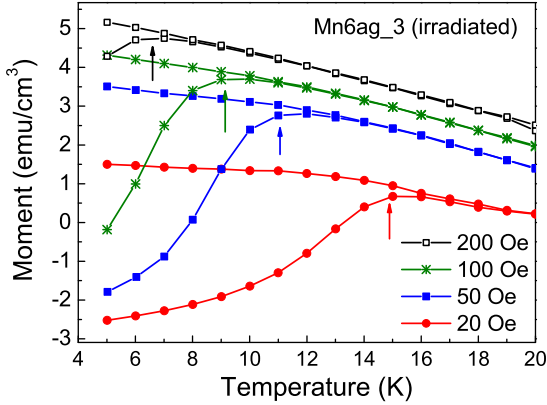}
\caption{ZFC/FC magnetization curves for the sample Mn6ag after
irradiation. The measurements were performed in different fields.
The curves are vertically shifted for better visibility. The
disparity between the ZFC and FC curves is the
signature of superparamagnetism. The blocking temperatures
(indicated by arrows) are shifted to lower temperatures with
increasing measurement field.}\label{fig:Mn490_ZFCFC}
\end{figure}

The appearance of a superparamagnetic component is demonstrated by
measuring the zero-field-cooled/field-cooled (ZFC/FC)
temperature-dependent magnetization for sample Mn6ag\_3, which is heavily
compensated but still shows magnetic hysteresis. In the ZFC measurement,
the sample was cooled down from around 300\,K to 5\,K in zero field.
Then a small field (in this work 20, 50, 100, 200\,Oe,
respectively) was applied. The magnetization was measured during
warming up to 300\,K. Then the field was kept and the sample was
cooled from 300\,K to 5\,K. During cooling, the FC magnetization was
recorded. A similar approach was used in Ref.\
\onlinecite{Sawichi2010}. If there is a superparamagnetic
component, one expects to observe a difference between the
ZFC and FC curves, as is indeed found in Fig.\ \ref{fig:Mn490_ZFCFC}. The
maximum in ZFC magnetization occurs at the so-called blocking
temperature, which depends on the size of superparamagnetic
particles and on the external field. For larger external field,
the single-domain superparamagnetic clusters can flip
into the field direction at lower temperature. In our
measurement, the blocking temperature decreases as the external
field is increased from 20\,Oe to 200\,Oe.

\begin{figure} \center
\includegraphics[scale=0.3]{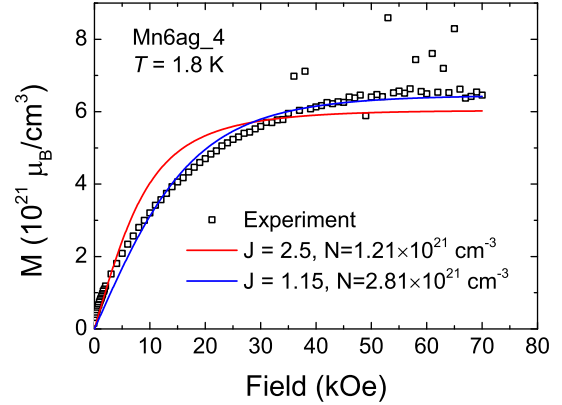}
\caption{Magnetization measured at 1.8\,K for sample Mn6ag\_4 (the most
highly compensated sample). The diamagnetic background has been subtracted.
The solid lines are fits using the Brillouin function. The fitting results are
also shown in the figure.}\label{fig:Mn490_paramagnetism}
\end{figure}

Figure \ref{fig:Mn490_paramagnetism} shows the magnetization for sample
Mn6ag\_4 measured at 1.8\,K after subtracting the diamagnetic background. This
sample does not show any ferromagnetic hysteresis down to 1.8 K. At 1.8\,K, the
magnetization does not show saturation up to a field of 7\,T. We fit the
magnetization by a Brillouin function,
\begin{eqnarray}
 M(\alpha) &=& NJ{\mu_B}g\,
  \bigg[ \frac{2J+1}{2J}\, \coth\left(\frac{2J+1}{2J}\, \alpha\right)
  \nonumber \\
&& {}- \frac{1}{2J}\, \coth\left(\dfrac{\alpha}{2J}\right) \bigg],
\label{Brillouin}
\end{eqnarray}%
where the $g$ factor is about 2 if assuming Mn$^{2+}$ ($d^5$) without holes
\cite{schneider1987electronic,pawlowski2006mn}, $\mu_B$ is the Bohr magneton,
$\alpha=gJ\mu_BH/k_BT$, $k_B$ is the Boltzmann constant, and $N$ is the
density of spins.

As shown in Fig.\ \ref{fig:Mn490_paramagnetism}, the fit for fixed
$J = 2.5$ does not well reproduce the experimental data. A
better fit is obtained for $J = 1.15$. This can be
explained by considering another component of paramagnetism arising from
intrinsic defects in the GaAs substrate \cite{PhysRevLett.52.851}, which technically cannot be subtracted
from the magnetization measurement since the defect concentration can be different depending on the growth condition. This component can be represented by spin
$1/2$ ($J = S = 0.5$) paramagnetism. Moreover, the other states of Mn impurities could have different $g$ factor and $J$ value \cite{pawlowski2006mn}. Therefore, it is difficult to accurately separate different paramagnetic components and is beyond the scope of this paper. Nevertheless, a fully compensated (Ga,Mn)As sample shows paramagnetic behavior.

\section{Hole concentration}
\label{app.b}

\begin{figure} \center
\includegraphics[scale=0.3]{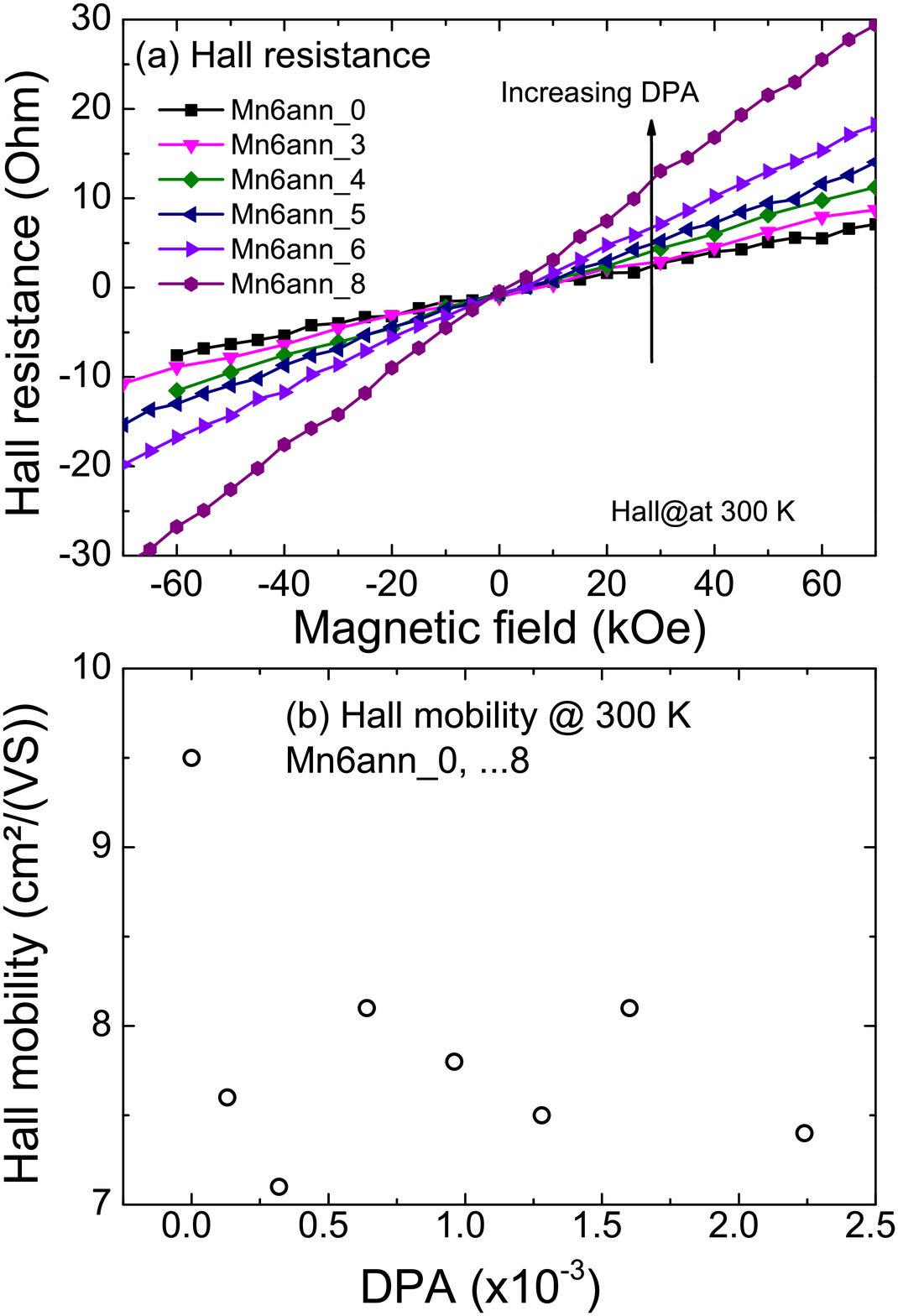}
\caption{(a) Hall resistance vs.\ applied field measured for several samples
from Mn6ann after irradiation and (b) Hall mobility for the same set of samples. All samples show p-type
conductivity and the hole concentration decreases gradually. The Hall mobility
decreases after irradiation but remains rather independent of DPA.}\label{fig:Mn377_Hall300K}
%with different Hall coefficients.
\end{figure}

Figure \ref{fig:Mn377_Hall300K}(a) shows the Hall resistance measured
at 300\,K for sample Mn6ann. For increasing DPA, the samples remain
p-type conducting, but the Hall
resistance increases, indicating a decrease of the hole
concentration. Estimated from the Hall measurement shown in
Fig.\ \ref{fig:Mn377_Hall300K}(a), the hole concentration gradually
decreases from around $2.6 \times 10^{20}\,\mathrm{cm}^{-3}$ for the virgin
sample to $5.6 \times 10^{19}\,\mathrm{cm}^{-3}$ for sample
Mn6ann\_8. However, due to the paramagnetic
component in the Hall resistance \cite{wang2012studies}, the hole
concentration is significantly underestimated. We have also estimated
the Hall mobility, which decreases after irradiation but remains rather
independent of DPA, see Fig.\ \ref{fig:Mn377_Hall300K}(b).
However, one has to note the possible overestimation of the Hall
mobility due to the underestimated hole concentration.

%\blue{However}, Mayer
%\textit{et al.}\ \blue{\cite{PhysRevB.81.045205} have}
%used electrochemical capacitance voltage profiling to
%measure the hole concentration and confirmed the linear
%correlation between hole concentration and DPA.

\begin{figure} \center
\includegraphics[scale=0.3]{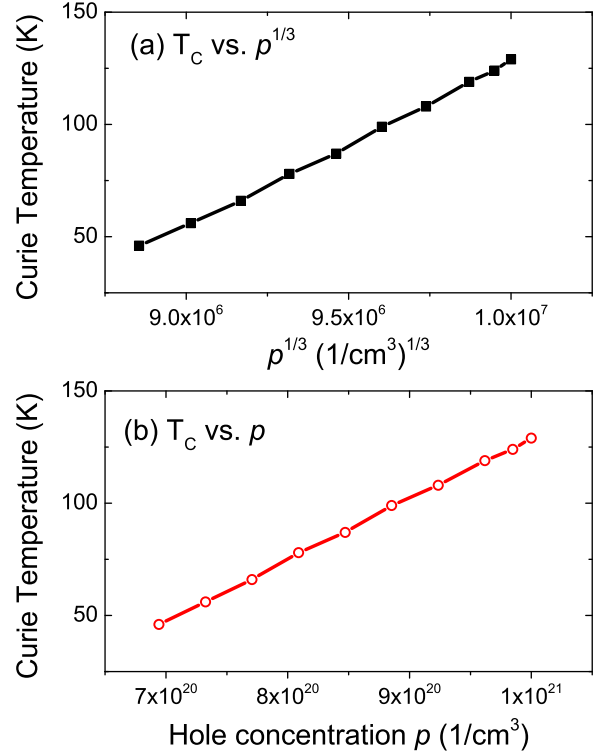}
\caption{Curie temperature vs. hole concentration $p$: (a) $p^{1/3}$ and (b) $p$ for samples Mn6ann$\_$0, …9. The hole concentration is estimated by following the approach in Ref. \onlinecite{PhysRevB.81.045205}. }\label{fig:Mn377_Tcp}
%with different Hall coefficients.
\end{figure}

The initial hole concentration is around $1 \times 10^{21}\,\mathrm{cm}^{-3}$, which has been measured by the sample preparation group at low temperature and high field on a similar sample \cite{wang2012studies}. This value can also be confirmed from our PIXE/channeling measurement. If we assume the nonrandom Mn fraction as the substitutional fraction (75\%) and most compensating Mn interstials have been kicked out during the postgrowth, long-time low-temperature annealing, the hole concentration is $2.21 \times 10^{22}\,\mathrm{cm}^{-3} \times  0.06  \times  0.75 = 0.99 \times 10^{21}\,\mathrm{cm}^{-3}$. The hole compensation rate by ion irradiation is estimated to be 0.91 hole per vacancy \cite{PhysRevB.81.045205}. Then we can calculate the hole concentration $p$ for sample Mn6ann after irradiation and plot $T_C$ vs. $p^{1/3}$ as shown in Figure \ref{fig:Mn377_Tcp}. As one can see, $T_C$ reveals a linear dependence on $p^{1/3}$. However, the dependence is not sensitive to the exponent 1/3 or 1 in the range of the hole concentration of our samples. Note that the exponent 1/3 only follows for a parabolic dispersion and neglecting disorder \cite{PhysRevB.72.165204}. Our argument does not rely on the specific value of the exponent but on the monotonicity.

\end{document}